\documentclass[12pt,prd,nofootinbib,superscriptaddress]{revtex4}

\usepackage{amsmath}
\usepackage{amsfonts}

\begin{document}
\title{Not on but of.}
\author{Olaf DREYER}
\affiliation{Dipartimento di Fisica, Universit\`a di Roma ``La Sapienza"\\
and Sez.~Roma1 INFN, P.le A. Moro 2, 00185 Roma, Italy}

\begin{abstract}  In physics we encounter particles in one of two ways.  Either as fundamental constituents of the theory or as emergent excitations.  These two ways differ by how the particle relates to the background.  It either sits \emph{on} the background,  or it is an excitation \emph{of} the background.  We argue that by choosing the former to construct our fundamental theories we have made a costly mistake.  Instead we should think of particles as excitations of a background.  We show that this point of view sheds new light on the cosmological constant problem and even leads to observable consequences by giving a natural explanation for the appearance of MOND-like behavior.  In this context it also becomes clear why there are numerical coincidences between the MOND acceleration  parameter $a_0$,  the cosmological constant $\Lambda$ and the Hubble parameter $H_0$. 
\end{abstract}

\maketitle

\tableofcontents

\newpage

\section{Which of our basic physical assumptions are wrong?}\label{sec:intro}
In theoretical physics we encounter particles in two different ways.  We either encounter them as fundamental constituents of a theory or as emergent entities.  A good example of the first kind is a scalar field.  Its dynamics is given by the Lagrangian
\begin{equation}
\int d^4x \left( \frac{1}{2}(\partial\phi)^2 - m^2\phi^2 + \ldots \right).
\end{equation}
An example of an emergent particle is a spin-wave in a spin-chain.  It given by
\begin{equation}
\vert k \rangle = \sum_{n=1}^{N} \exp{\left(2\pi i \frac{n k}{N}\right)} \vert n \rangle
\end{equation}
If $\vert 0 \rangle$ is the gound state of the spin-chain then $\vert n \rangle$ is the state where spin $n$ is flipped.  

Of all the differences between these two concepts of particles we want to stress one in particular:  the relation these particles have to their respective backgrounds.

The scalar field is formulated as a field \emph{on} spacetime.  The above Lagrangian does not include the metric.  If we include the metric we obtain
\begin{equation}
\int d^4x \sqrt{-g} \left( \frac{1}{2}g^{ab}\partial_a\phi\partial_b\phi - m^2\phi^2 + \ldots \right). 
\end{equation}

This way the scalar field knows about a non-trivial spacetime.  Einstein told us that this is not a one-way street.  The presence of a scalar field in turn changes the background.  The proper equations that describe this interaction between matter and background are obtained by adding the curvature tensor to the above Lagrangian.  This interaction does not change the basic fact about the scalar field that it is sitting \emph{on} spacetime.  It is distinct from the spacetime that it sits on.

This is in contrast to the emergent particle.  There is no clean separation possible between the particle and the background, i.e. the ground state in this case.  The spin-wave above is an excitation \emph{of} the background not an excitation \emph{on} the background.

We can now state what basic assumption we think needs changing.  Currently we built our fundamental theories assuming the matter-on-background model.  We will argue here that this assumption is wrong and that instead all matter is of the emergent type and that the excitation-of-background model applies.

\begin{description}
\item[Our basic assumption that is wrong:]  Matter does not sit \emph{on} a background but is an excitation \emph{of} the background.  
\end{description}

In the following we will do two things.  First we will show that the assumption that matter sits on a background creates problems and how our new assumption avoids these problems.  Then we will argue that the new assumption has observable consequences by showing how MOND like behavior arises naturally.

\section{The cosmological constant}
The basic assumption that matter sits on a background directly leads to one of the thorniest problems of theoretical physics:  the cosmological constant problem.  All matter is described by quantum fields which can be thought of as a collection of quantum harmonic oscillators.  One feature of the spectrum of the harmonic oscillator is that it has a non-vanishing ground state energy of $\hbar\omega/2$.  Because the quantum field sits on spacetime the ground state energy of all the harmonic oscillators making up the field should contribute to the curvature of space.  The problem with this reasoning is that it leads to a prediction that is many orders of magnitude off.  In fact if one assumes that there is some large frequency limit $\omega_\infty$ for the quantum field then the energy density coming from this ground state energy is proportional to the fourth power of this frequency:

\begin{equation}
\epsilon  \simeq  \omega_\infty^4
\end{equation}

If one chooses $\omega_\infty$ to be the Planck frequency then one obtains a value that is 123 orders of magnitude larger than the observed value of the cosmological constant.  This has been called the worst prediction of theoretical physics and constitutes one part of the cosmological constant problem (see \cite{Weinberg:1989p695,Carroll:2001p311} for more details).

Let us emphasize again how crucial the basic assumption of matter-on-background is for the formulation of the cosmological constant problem.  Because we have separated matter from the background we have to consider the contributions coming from the ground state energy.  We have to do this even when the matter is in its ground state,  i.e. when no particles are present.  This is to be contrasted with the excitation-of-background model.  If there are no waves present there is also no ground state energy for the excitations to consider.  The cosmological constant problem can not be formulated in this model.

There are two objections to this reasoning that we have to deal with.  The first objection is that although the above reasoning is correct it is also not that interesting because there is no gravity in the picture yet.  We will deal with this objection in the next section.  The other objection concerns the ontological status of the vacuum fluctuations.  IsnÕt it true that we have observed the Casimir effect?  Since the Casimir effect is based on vacuum fluctuations the cosmological constant problem is not really a problem that is rooted in any fundamental assumption but in observational facts.  For its formulation no theoretical foundation needs to be mentioned.  We will deal with this objection now.

There are two complementary views of the Casimir effect and the reality of vacuum fluctuations.  In the more well known explanation of the Casimir force between two conducting plates the presence of vacuum fluctuations is assumed.  The Casimir force then arises because only a discrete set of frequencies is admissible between the plates whereas no such restriction exists on the outside.  The Casimir effect between two parallel plates has been observed and this has led to claims that we have indeed seen vacuum fluctuations.  This claim is not quite true because there is another way to think about the Casimir effect.  In this view vacuum fluctuations play no role.  Instead,  the Casimir effect results from forces between the currents and charges in the plates.  No vacuum diagrams contribute to the total force\cite{Jaffe:2005p147}.  If the emergent matter is described by the correct theory,  say quantum electrodynamics,  we will find the Casimir effect even if there are no vacuum fluctuations.  

We see that the cosmological constant problem can be seen as a consequence of us viewing matter as sitting on a background.  If we drop this assumption we can not even formulate the cosmological constant problem.

\section{Gravity}\label{sec:gravity}
The above argument for viewing matter as an excitation of a background is only useful if we can include gravity in the picture.  In \cite{dreyer} we have argued that this can be achieved by regarding the ground state itself as the variable quantity that is responsible for gravity.  In the simplest case the vacuum is described by a scalar quantity $\theta$.  If we assume that the energy of the vacuum is given by 
\begin{equation}\label{eqn:hamiltonian}
E = \frac{1}{8\pi}\int d^3x\; (\nabla \theta)^2,
\end{equation}
then we can calculate the force between two objects $\mathsf{m}_i$, $i=1,2$.  If we  introduce the gravitational mass of an object by 
\begin{equation}\label{eqn:gravitational}
\mathsf{m} = \frac{1}{4\pi}\int_{\partial \mathsf{m}} d\sigma \cdot \nabla\theta,
\end{equation}
where $\partial \mathsf{m}$ is the boundary of the object $\mathsf{m}$,  then the force between them is given by 
\begin{equation}
F = \frac{\mathsf{m}_1\mathsf{m}_2}{r^2}.
\end{equation}
Usually we express Newton's law of gravitation not in terms of the gravitational masses $\mathsf{m}_i$, $i=1,2$,  but in terms of the inertial masses $m_i$,  $i=1,2$.  In \cite{dreyer} we have argued that the inertial mass of an object is given by
\begin{equation}\label{eqn:inertial}
m = \frac{2 \mathsf{m}}{3 a}\; \mathsf{m},
\end{equation}
where $a$ is the radius of the object.  In terms of the inertial masses $m_i$ Newton's law then takes the usual form
\begin{equation}
F = G \frac{m_1m_2}{r^2},
\end{equation}
where $G$ has to be calculated from the fundamental theory:
\begin{equation}
G = \left(\frac{3 a}{\mathsf{2 m}}\right)^2
\end{equation}
In this picture of gravity the metric does not play a fundamental role.  Gravity appears because the ground state $\theta$ depends on the matter.

\section{MOND as a consequence}
The picture of gravity that we have given in the last section is valid only for zero temperature.  If the temperature is not zero we need to take the effects of entropy into account and instead of looking at the energy we have to look at the free energy
\begin{equation}
F = E - TS.
\end{equation}
We thus have to determine the dependence of the entropy $S$ on the temperature $T$ and the ground state $\theta$.  The entropy should not depend on $\theta$ directly because every $\theta$ corresponds to the same ground state.  The entropy should only be dependent on changes of $\theta$. If we are only interested in small values of $T$ we find 
\begin{equation}
S = \sigma T(\nabla\theta)^2,
\end{equation}
for some constant $\sigma$.  The total free energy is thus
\begin{eqnarray}
F & = & E - TS \\
  & = & E (1 - 8 \pi \sigma T^2). 
\end{eqnarray}
We see that a non-zero temperature does not change the form of the force but just its strength.  The new gravitational constant is given by
\begin{equation}
G_{T}  =  \frac{1}{1 - \sigma T^2} \ G_{T=0}.
\end{equation}
The situation changes in an interesting way if there is a large maximal length scale $L_\text{\scriptsize{max}}$ present in the problem.  The contributions to the entropy of the form $(\nabla\phi)^2$ come from excitations with a wavelength of the order
\begin{equation}
L = \vert\nabla\phi\vert^{-1}.
\end{equation}
If this wavelength $L$ is larger than $L_\text{\scriptsize{max}}$ than these excitations should not exist and thus not contribute to the entropy.  Instead of a simple $(\nabla\phi)^2$ term we should thus have a term of the form
\begin{equation}
C( L_\text{\scriptsize{max}}\; \vert\nabla\theta\vert ) \cdot (\nabla\theta)^2,
\end{equation}
where the function $C$ is such that it suppresses contributions from excitations with wavelengths larger than $L_\text{\scriptsize{max}}$.  For wavelengths much smaller than the maximal wavelength we want to recover the usual contributions to the entropy.  Thus,  if $L_\text{\scriptsize{max}}\cdot\vert\nabla\theta\vert$ is much bigger than unity we want $C$ to be one:
\begin{equation}
C(x) = 1, \text{\ \ \ for\ \ \ } x \gg 1.
\end{equation}
For $x\ll 1$ we assume that the function $C$ possesses a series expansion of the form
\begin{equation}
C(x) = \alpha x + \beta x^2 + \ldots.
\end{equation}
For small values of $L_\text{\scriptsize{max}}\cdot\vert\nabla\theta\vert$ we thus find that the dependence of the entropy on $\phi$ is of the form
\begin{equation}
T^2 \sigma \int d^4x\; \alpha L_\text{\scriptsize{max}}\vert\nabla\phi\vert^3.
\end{equation}
It is here that we make contact with the Lagrangian formulation of Milgrom's odd new dynamics (or MOND,  see \cite{Milgrom:1983p708} and \cite{Famaey:2011p753} for more details).  In \cite{Bekenstein:1984p709} Bekenstein and Milgrom have shown that a Lagrangian of the form
\begin{equation}
\int d^3x \left( \rho\theta + \frac{1}{8\pi G}\; a_0^2\; F\left[ \frac{(\nabla\theta)^2}{a_0^2} \right]\right)
\end{equation}
gives rise to MOND like dynamics if the function $F$ is chosen such that 
\begin{equation}
\mu(x) = F^\prime(x^2). 
\end{equation}
Here $\mu(x)$ is the function that determines the transition from the classical Newtonian regime to the MOND regime.  It satisfies
\begin{equation}
\mu(x) = \left\{ %
\begin{array}{cl}
	1	&	x\gg 1 \\
	x	& 	x\ll 1
\end{array}\right.
\end{equation}
From this it follows that the function $F$ satisfies
\begin{equation}
F(x^2) = \left\{ %
\begin{array}{cl}
	x^2			&	x\gg 1 \\
	2/3\;x^3	& 	x\ll 1
\end{array}\right.
\end{equation}
The behavior of the Lagrangian is then
\begin{equation}\label{eqn:mondlagrangian}
a_0^2 F\left( \frac{\vert\nabla\theta\vert^2}{a_0^2} \right) = \left\{ %
\begin{array}{cl}
		\vert\nabla\theta\vert^2	&   \ \ \ a_0^{-1} \vert\nabla\theta\vert \gg 1 \\
	\frac{2}{3 a_0}\;\vert\nabla\theta\vert^3	& 	\ \ \ a_0^{-1} \vert\nabla\theta\vert \ll 1
\end{array}\right.
\end{equation}
This is exactly the behavior of the free energy that we have just derived if we make the identification
\begin{equation}\label{eqn:eqn}
\frac{2}{3 a_0} = \alpha \sigma T^2 L_\text{\scriptsize{max}}.
\end{equation}
There are currently two candidates for a maximal length scale $L_\text{\scriptsize{max}}$.  These are the Hubble scale
\begin{equation}
L_H = c H_0
\end{equation}
and the cosmic horizon scale
\begin{equation}
L_\Lambda = \sqrt{\frac{1}{\Lambda}}.
\end{equation}
It is a remarkable fact of the universe that we live in that \emph{both} of these length scales satisfy the relationship that we derived in (\ref{eqn:eqn}) if we further assume that the constant $\alpha \sigma T^2$ is of order one.  We have thus established a connection between the acceleration parameter $a_0$,  the cosmological constant $\Lambda$,  and the Hubble parameter $H_0$.  In standard cosmology these coincidences remain complete mysteries. 

\section{Discussion}
Particles are either fundamental or they are emergent.  If they are fundamental they are sitting on a background;  if they are emergent they are excitations of a background.  Rather than being a purely philosophical issue we have argued that this distinction is important and that the assumption that particles are fundamental is wrong.  Assuming instead that particles are emergent leads to the resolution of theoretical problems as well as having observational consequences.  We have argued that the cosmological constant problem as it is usually formulated can not even be stated if we think of particles as excitations of a background.  Also,  we have shown that this picture gives a straight forward way of understanding the appearance of MOND like behavior in gravity.  The argument also makes clear why there are numerical relations between the MOND parameter $a_0$,  the cosmological constant $\Lambda$,  and the Hubble parameter $H_0$.

Our derivation of MOND is inspired by recent work \cite{Ho:2011p703,Ho:2010p711,Pikhitsa:2010p769,Klinkhamer:2012p750,Klinkhamer:2011p756,Li:2011p725,Neto:2011p723,Kiselev:2010p717,Pazy:2012p789,Modesto:2010p702} that uses Verlinde's derivation of Newton's law of gravity \cite{verlinde} as a starting point.  Our derivation differs from these in that it does not rely on holography in any way.  Our formulae for the entropy are all completely three-dimensional.

\section*{Acknowledgements}
I would like to thank Seth Llloyd,  Mamazim Kassankogno, and Stevi Roussel Tankio Djiokap for helpful discussion and the Foundational Questions Institute, FQXi, for financial support and for creating an environment where it is alright to play with unorthodox ideas.


\begin{thebibliography}{WW}
\bibitem{Weinberg:1989p695}  S.~Weinberg. \emph{The cosmological constant problem.} Reviews of Modern Physics (1989) vol. 61 (1) pp. 1-23.
\bibitem{Carroll:2001p311}Carroll. \emph{The cosmological constant.} Living Rev. Relativ. (2001) vol. 4 pp. 2001-1, 80 pp. (electronic).
\bibitem{Jaffe:2005p147} R.~L.~Jaffe.  \emph{Casimir effect and the quantum vacuum.} Phys. Rev. D (2005) vol. 72 (2) pp. 5.
\bibitem{dreyer} O.~Dreyer,  \emph{Internal relativity},  arXiv:1203.2641.
\bibitem{Milgrom:1983p708} Milgrom. \emph{A modification of the Newtonian dynamics as a possible alternative to the hidden mass hypothesis.} The Astrophysical Journal (1983) vol. 270 pp. 365-370.
\bibitem{Famaey:2011p753}  Famaey and McGaugh. \emph{Modified Newtonian Dynamics (MOND): Observational Phenomenology and Relativistic Extensions.} Arxiv preprint arXiv:1112.3960v2 (2011).
\bibitem{Bekenstein:1984p709} Bekenstein and Milgrom. \emph{Does the missing mass problem signal the breakdown of Newtonian gravity? } The Astrophysical Journal (1984) vol. 286 pp. 7-14.
\bibitem{Ho:2011p703} Ho et al. \emph{Quantum gravity and dark matter.} Gen Relativ Gravit (2011) vol. 43 (10) pp. 2567-2573.
\bibitem{Ho:2010p711} Ho et al. \emph{Cold dark matter with MOND scaling.} Physics Letters B (2010) vol. 693 (5) pp. 567-570.
\bibitem{Pikhitsa:2010p769} Pikhitsa. \emph{MOND reveals the thermodynamics of gravity.} Arxiv preprint arXiv:1010.0318 (2010).
\bibitem{Klinkhamer:2012p750} Klinkhamer. \emph{Entropic-Gravity Derivation of MOND.} Modern Physics Letters A (2012).
\bibitem {Klinkhamer:2011p756} Klinkhamer and Kopp. \emph{Entropic gravity, minimum temperature, and modified Newtonian dynamics.} Arxiv preprint arXiv:1104.2022v6 (2011).
\bibitem{Li:2011p725} Li and Chang. \emph{Debye Entropic Force and Modified Newtonian Dynamics}. Communications in Theoretical Physics (2011) vol. 55 pp. 733-736.
\bibitem{Neto:2011p723} Neto. \emph{Nonhomogeneous Cooling, Entropic Gravity and MOND Theory.} Internat. J. Theoret. Phys. (2011) vol. 50 pp. 3552-3559.
\bibitem{Kiselev:2010p717} Kiselev and Timofeev. \emph{The holographic screen at low temperatures.} arXiv (2010) Arxiv preprint arXiv:1009.1301v2 (2010).
\bibitem{Pazy:2012p789} Pazy and Argaman. \emph{Quantum particle statistics on the holographic screen leads to modified Newtonian dynamics.} Physical Review D (2012) vol. 85 (10) pp. 104021.
\bibitem{Modesto:2010p702} Modesto and Randono. \emph{Entropic corrections to Newton's law.} arXiv (2010)Arxiv preprint arXiv:1003.1998v1 (2010).
\bibitem{verlinde} E. Verlinde, \emph{On the Origin of Gravity and the Laws of Newton},  	JHEP 1104:029,  2011.
\end{thebibliography}
\end{document}